\documentclass[aps,prd,showpacs,preprintnumbers,twocolumn,nofootinbib]{revtex4}
\usepackage[dvips]{graphicx}
\usepackage{bm,latexsym,amsmath,amssymb,amsfonts}
\newcommand*{\cP}{{\cal P}}
\newcommand*{\mpl}{M_{{\rm Pl}}}
\begin{document}

\title{
Initial Kaluza-Klein fluctuations and inflationary gravitational waves
in braneworld cosmology
}

\author{Tsutomu~Kobayashi}
\email{tsutomu@tap.scphys.kyoto-u.ac.jp}

\affiliation{
Department of Physics, Kyoto University, Kyoto 606-8502, Japan 
}

\begin{abstract}
We study the spectrum of gravitational waves generated from inflation
in the Randall-Sundrum braneworld.
Since the inflationary gravitational waves are of quantum-mechanical origin,
the initial configuration of perturbations in the bulk includes
Kaluza-Klein quantum fluctuations as well as fluctuations in the zero mode.
We show, however, that the initial fluctuations in Kaluza-Klein modes
have no significant effect on the late time spectrum,
irrespective of
the energy scale of inflation and the equation of state parameter
in the post-inflationary stage.
This is done numerically, using the Wronskian formulation.
\end{abstract}


\preprint{KUNS-2011}

\maketitle
%%%%%%%%%%%%%%%%%%%%%%%%%%%%%%%%%%%%%%%%%%%%%%%%%%%%%%%%%%%%%%%%%%%%%%

\section{Introduction}

The inflationary scenario predicts the existence
of a gravitational wave background produced
quantum-mechanically~\cite{Maggiore:1999vm},
in much the same way as the mechanism of generating
density perturbations.
Detecting the inflationary gravitational wave background
is a  grand challenge for LISA~\cite{LISA},
BBO~\cite{BBO}, and other missions~\cite{DECIGO}.
It will open up a new window into
the very early universe,
and thus in principle the possibility
of recently proposed braneworld models~\cite{Maartens:2003tw}
can be tested using the inflationary gravitational waves.

%\cite{Tanaka:2004ig, Kobayashi:2004wy}
%\cite{Langlois:2000ns}
%\cite{Kobayashi:2003cn}
%\cite{Kobayashi:2005jx}
%\cite{Tanaka:2004ig}
%\cite{Kobayashi:2004wy}
%\cite{Hiramatsu:2003iz}
%\cite{Hiramatsu:2004aa}
%\cite{Ichiki:2003hf}
%\cite{Ichiki:2004sx}
%\cite{Hiramatsu:2006bd}
%\cite{Kobayashi:2005dd}
%\cite{Gorbunov:2001ge}
%\cite{Easther:2003re}
%\cite{Battye:2003ks}
%\cite{Battye:2004qw}
%\cite{Cartier:2005br}

%\cite{Langlois:2000ns,Kobayashi:2003cn,Kobayashi:2005jx,Tanaka:2004ig,Kobayashi:2004wy,Hiramatsu:2003iz,Hiramatsu:2004aa,Ichiki:2003hf,Ichiki:2004sx,Hiramatsu:2006bd,Kobayashi:2005dd,Gorbunov:2001ge,Easther:2003re,Battye:2003ks,Battye:2004qw,Cartier:2005br}

A number of studies have been done
aiming at obtaining a clear picture
of the generation and evolution of gravitational waves
in the Randall-Sundrum braneworld~\cite{Langlois:2000ns,Kobayashi:2003cn,Kobayashi:2005jx,Tanaka:2004ig,Koyama:2004cf,Kobayashi:2004wy,Hiramatsu:2003iz,Hiramatsu:2004aa,Ichiki:2003hf,Ichiki:2004sx,Hiramatsu:2006bd,Kobayashi:2005dd,Gorbunov:2001ge,Easther:2003re,Battye:2003ks,Battye:2004qw,Cartier:2005br}.
De Sitter inflation on the brane~\cite{Langlois:2000ns} is a special case
where the perturbation equation
is separable and exactly solvable.
It predicts a flat primordial spectrum, as in four-dimensional general relativity,
but the amplitude is enhanced at high energies.
For more general inflation models, the tilted primordial spectrum can be
calculated from the corresponding four-dimensional one
by a mapping formula~\cite{Kobayashi:2003cn, Kobayashi:2005jx}.
The late time evolution of gravitational wave perturbations
at low energies
is very close to that in conventional
cosmology~\cite{Tanaka:2004ig, Koyama:2004cf, Kobayashi:2004wy},
recovering
four-dimensional general relativity
when the relevant length scales are lager than the bulk curvature scale $\ell$.

What is much more interesting is
the evolution of gravitational waves during the high energy regime,
$\ell H\gg 1$, where $H$ is the Hubble parameter.
In addition to the rather trivial effect of the unconventional
background expansion rate,
the nontrivial effect of mode mixing may not be negligible,
converting a zero mode into a Kaluza-Klein (KK) mode
and vice versa efficiently
at high energies.
To clarify the evolution of subhorizon modes
during the high energy regime,
several numerical studies have been done
with different numerical schemes~\cite{Hiramatsu:2003iz,Hiramatsu:2004aa,Ichiki:2003hf,Ichiki:2004sx,Hiramatsu:2006bd}.
Most of them concentrated on the radiation-dominated phase
after inflation,
and very recently Hiramatsu~\cite{Hiramatsu:2006bd} investigated
consequences of a different equation of state parameter $w$.
All of the above papers consider very na\"{i}ve initial conditions,
ignoring initial abundance of KK fluctuations.
However, since the gravitational wave perturbations
are of quantum-mechanical origin,
the most plausible initial configuration of the perturbations during inflation
should include vacuum fluctuations in KK modes.
Taking this point into account,
in Refs.~\cite{Kobayashi:2005jx, Kobayashi:2005dd}
Kobayashi and Tanaka have developed
a numerical formulation using the Wronskian,
by which one can discuss the quantum-mechanical generation
and subsequent evolution of gravitational waves.
Following the previous works~\cite{Kobayashi:2005jx, Kobayashi:2005dd},
in the present paper we revisit the spectrum of
inflationary gravitational wave background in the braneworld,
focusing in particular on the effect of initial abundance of the KK fluctuations.

The paper is organized as follows.
In the next section we describe the background cosmological model,
and then in Sec.~III we review the Wronskian formulation
to compute the spectrum of inflationary gravitational waves numerically.
We present our numerical results in Sec.~IV.
Our conclusions are drawn in Sec.~V.

\section{The background model}

We will work in the Randall-Sundrum-type braneworld~\cite{Randall:1999vf},
and the bulk metric is given by the five-dimensional
anti-de Sitter spacetime
\begin{eqnarray}
ds^2=\frac{\ell^2}{z^2}\left(-dt^2+\delta_{ij}dx^idx^j+dz^2\right),
\end{eqnarray}
where $\ell$ is the bulk curvature scale.
A cosmological solution can be described by
a moving brane in the above static coordinates.
The scale factor $a(t)$ can be expressed in terms of
the location of the brane $z(t)$ as $a(t)=\ell/z(t)$,
and it is governed by the modified Friedmann equation~\cite{FRW_brane}
\begin{eqnarray}
H^2=\frac{\rho}{3\mpl^2}\left(1+\frac{\rho}{2\sigma}\right),
\label{friedmanneq}
\end{eqnarray}
where $\rho$ is the matter energy density on the brane and
$\sigma=6\mpl^2/\ell^2$ is the tension of the brane.
The matter content is characterized by the
equation of state parameter
\begin{eqnarray}
w:=\frac{p}{\rho}.
\end{eqnarray}
Since the standard conservation law
$d\rho/d\tau=-3H(\rho+p)$
holds on the brane,
we have $\rho\propto a^{-3(1+w)}$ when $w$ is constant.
Using this and Eq.~(\ref{friedmanneq}) we obtain 
the scale factor as a function of the proper time $\tau$ on the brane.

The initial stage of the background model is
assumed to be described by a de Sitter brane with a constant expansion rate $H_i$,
and the final stage by a Minkowski brane.
The two stages are connected smoothly by
a Friedmann-Robertson-Walker (FRW) brane
with a constant equation of state parameter $w$.
We dub this phase as the ``FRW phase''.
The connection to the Minkowski phase is done at sufficiently low energies
because we would like to focus on the possible high energy effects
at the early stage of the FRW phase just after inflation.
The construction here allows us to
discuss the quantum-mechanical generation of gravitational waves
during inflation and their subsequent evolution during the FRW phase,
and to see the final amplitude of the well-defined zero mode.
In the previous paper~\cite{Kobayashi:2005dd} only
the radiation-dominated stage ($w=1/3$) was considered, but
in the present paper we are interested in other various values of $w$ as well.

\section{Formulation}

\subsection{Double null coordinates}

Let us consider gravitational wave (tensor-type) perturbations.
The perturbed metric can be written as
\begin{eqnarray}
ds^2=\frac{\ell^2}{z^2}\left[-dt^2
+(\delta_{ij}+h_{ij})dx^idx^j+dz^2\right],
\end{eqnarray}
where $h_{ij}=h_{ij}(t, \textbf{x} ,z)$ is the transverse-traceless metric perturbation.
We decompose it into the spatial Fourier modes as
\begin{eqnarray}
h_{ij}=\frac{\sqrt{2}}{(2\pi M_5)^{3/2}}\int \!d^3k\phi_{\textbf{k}}(t,z)
e^{i\textbf{k}\cdot\textbf{x}}e_{ij},
\label{fourier}
\end{eqnarray}
where $M_5$ is the fundamental mass scale which is related to the
four-dimensional Planck mass $\mpl$ by $\ell(M_5)^3=\mpl^2$.
The linearized Einstein equations give
the Klein-Gordon-type equation for $\phi_{\textbf{k}}$:
\begin{eqnarray}
\left(\frac{\partial^2}{\partial t^2}+k^2-\frac{\partial^2}{\partial z^2}
+\frac{3}{z}\frac{\partial}{\partial z}\right)\phi_{\textbf{k}}=0,
\label{KGintz}
\end{eqnarray}
and the junction conditions at the brane read
\begin{eqnarray}
\left.n^a\partial_a\phi_{\textbf{k}}\right|_{{\rm brane}}=0,
\end{eqnarray}
where $n^a$ is the unit normal to the brane.
From now on we suppress the subscript $\textbf{k}$.

Following the previous papers~\cite{Kobayashi:2005jx, Kobayashi:2005dd}
we use the Wronskian formulation to compute the spectrum of
gravitational waves.
Basic details of the formulation and numerical scheme
are explained in Ref.~\cite{Kobayashi:2005jx},
but here we make a slight improvement.
Since double null coordinates are convenient for numerical studies,
our starting point is
\begin{eqnarray}
u=t-z,\quad v=t+z.
\end{eqnarray}
In this coordinate system, the trajectory of the brane can be specified
arbitrarily by
\begin{eqnarray}
v=q(u).
\end{eqnarray}
To simplify this boundary trajectory, we make a coordinate transformation
\begin{eqnarray}
U=u,\quad q(V)=v,
\end{eqnarray}
and then the position of the brane is given by
\begin{eqnarray}
U=V.
\end{eqnarray}
The following further coordinate transformation turns out to be useful
for actual numerical calculations:
\begin{eqnarray}
X&=&-\frac{\ell}{2}\ln\left(- \frac{U}{\ell}\right),
\\
T&=&-\frac{\ell}{2}\ln\left(- \frac{V}{\ell}\right).
\end{eqnarray}
Note that by an appropriate choice
of the origin of the time coordinate $t$, we can always have $U<0$ and $V<0$.
The position of the brane is simply given by
\begin{eqnarray}
T=X.
\end{eqnarray}
We will perform numerical computations with equal grid spacing $\Delta$
in both $T$ and $X$ directions.
The advantage of using these new coordinates is that
it is easier to extend the computational domain far away from the brane
in these coordinates than in the $(U, V)$ coordinates
which were used in the previous studies.
The step size $\varepsilon$ in the $U$ ($V$) coordinate increases with
increasing $|X|$ ($|T|$), but we will adjust $\Delta$
so that the maximum of $\varepsilon$
is not so different from the step size used in
Ref.~\cite{Kobayashi:2005dd}.
Thanks to this choice of $\Delta$,
the $(T, X)$ coordinates will not cause any problems
in the resolution of the numerical analysis.

Now the bulk metric is written as
\begin{eqnarray}
ds^2=A^2(T,X)
\left[-2e^{-2X/\ell}\tilde q'(T)dTdX+\delta_{ij}dx^idx^j\right],\cr
\end{eqnarray}
where
\begin{eqnarray}
A(T,X)&:=&\frac{2\ell}{\tilde q(T)+\ell e^{-2X/\ell}},
\\
\tilde q(T)&:=&q(-\ell e^{-2T/\ell}),
\end{eqnarray}
and a prime stands for a derivative with respect to the argument.
The proper time $\tau$ and the scale factor $a$ on the brane are given
respectively by
\begin{eqnarray}
d\tau &=&a\cdot \sqrt{2\tilde q'(T)}e^{-T/\ell} dT
\label{dtau<->dT},
\\
a&=&A(T, T),
\end{eqnarray}
and hence we have
\begin{eqnarray}
\tilde q'(T)=2e^{-2T/\ell}\left(\sqrt{1+\ell^2 H^2}-\ell H\right)^2.
\label{dq<->H}
\end{eqnarray}
Given the Hubble parameter $H$ as a function of $\tau$,
one can integrate Eqs.~(\ref{dtau<->dT}) and~(\ref{dq<->H})
to obtain $\tilde q$ as a function of $T$.

In the $(T, X)$ coordinates
the Klein-Gordon-type equation for $\phi$ reduces to
\begin{eqnarray}
&&2\partial_T\partial_X\phi +\frac{3}{\tilde q(T)+\ell e^{-2X/\ell}}
\left[2e^{-2X/\ell}\partial_T-\tilde q'(T)\partial_X\right]\phi
\nonumber\\
&&\qquad +e^{-2X/\ell}\tilde q'(T)k^2\phi =0,
\label{KGinTX}
\end{eqnarray}
subject to the boundary condition
\begin{eqnarray}
\left.\left[\partial_T-\partial_X\right]\phi\right|_{T=X}=0.
\end{eqnarray}

\subsection{Wronskian formulation}

Since the system we are considering has infinitely many degrees of freedom,
consisting of a zero mode and a tower of KK modes,
it is not advisable to set up the initial value problem
for the evolution of each perturbation modes.
Instead, we confine our attention to a single degree of freedom
concerning the zero mode in the final Minkowski phase.
The necessary information that determines the amplitude
of the final zero mode
can be picked up by using the Wronskian.
Our Wronskian evaluated on a constant $T$ hypersurface is defined by
\begin{eqnarray}
\left(\phi_1\cdot\phi_2\right) :=
2i\int_{-\infty}^{T}\!\!dXA^3(T, X)
\left(\phi_1\partial_X\phi_2^*-\phi_2^*\partial_X\phi_1\right),\cr
\end{eqnarray}
which is independent of the choice of the hypersurface.

Treating the graviton field $\phi$ as an operator,
in the initial de Sitter phase we expand it
using the annihilation and creation operators,
$\hat a_n$ and $\hat a_n^{\dagger}$, as
\begin{eqnarray}
\hat\phi = \hat a_0\phi_0+\hat a_0^{\dagger}\phi_0^*
+\int_0^{\infty}\! d\nu\left(\hat a_{\nu}\phi_{\nu}+\hat a_{\nu}^{\dagger}\phi_{\nu}^*\right),
\end{eqnarray}
where $\phi_n$ and its complex conjugate $\phi_n^*$
are the mode functions in the de Sitter phase
($n=0$ for the zero mode and $n=\nu$ for the KK modes).
These mode functions
form a complete orthonormal basis with respect to the Wronskian:
\begin{eqnarray}
&&(\phi_0 \cdot \phi_0) = -(\phi_0^*\cdot\phi_0^*) = 1,
\label{Wronskian_condition_dS}\\
&&(\phi_{\nu} \cdot \phi_{\nu'}) = - (\phi_{\nu}^*\cdot\phi_{\nu'}^*)
= \delta(\nu-\nu'),
\nonumber\\
&&(\phi_0 \cdot \phi_{\nu}) = (\phi_{0}^*\cdot\phi_{\nu}^*)
=0,
\nonumber\\
&&(\phi_{n}\cdot\phi_{n'}^*) = 0,
\qquad\mbox{for}\quad n,n'=0, \nu.
\nonumber
\end{eqnarray}
The index $\nu$ is related to the KK mass $m$ as
\begin{eqnarray}
m^2=\left(\nu^2+\frac{9}{4}\right)H_i^2,
\end{eqnarray}
and $\nu\geq 0$.

Similarly, 
in the final Minkowski phase we expand it as
\begin{eqnarray}
\hat\phi = \hat A_0\varphi_0+\hat A_0^{\dagger}\varphi_0^*
+\int_0^{\infty}\! dm\left(\hat A_m\varphi_m+\hat A_m^{\dagger}\varphi_m^*\right),
\end{eqnarray}
where $\varphi_n$ and $\varphi_n^*$ $(n=0,m)$
are the mode functions in the Minkowski phase,
and $\hat A_n$ and $\hat A_n^{\dagger}$ are
the annihilation and creation operators, respectively,
of their corresponding modes.
These mode functions also
form a complete orthonormal basis with respect to the Wronskian:
\begin{eqnarray}
&&(\varphi_0 \cdot \varphi_0) = -(\varphi_0^*\cdot\varphi_0^*) = 1,
\label{Wronskian_condition_min}\\
&&(\varphi_{m} \cdot \varphi_{m'}) = - (\varphi_{m}^*\cdot\varphi_{m'}^*)
= \delta(m-m'),
\nonumber\\
&&(\varphi_0 \cdot \varphi_{m}) = (\varphi_{0}^*\cdot\varphi_{m}^*)
=0,
\nonumber\\
&&(\varphi_{n}\cdot\varphi_{n'}^*) = 0,
\qquad\mbox{for}\quad n,n'=0, m,
\nonumber
\end{eqnarray}
The explicit form of the above mode functions is
presented in Appendix.

Using the Wronskian, we now explain how to compute
the spectrum of gravitational waves.
We assume that initially the gravitons are in the de Sitter invariant
vacuum state annihilated by $\hat a_0$ and $\hat a_{\nu}$,
\begin{eqnarray}
\hat a_0|0\rangle=\hat a_{\nu}|0\rangle=0.
\end{eqnarray}
The expectation value of the squared amplitude of
the zero mode in the final Minkowski stage is
\begin{eqnarray}
\langle 0|
\left( \hat A_0\varphi_0+\hat A_0^{\dagger}\varphi^*_0\right)^2
|0\rangle
\simeq\frac{1}{k\ell}N_f,\label{cP--Nf}
\end{eqnarray}
where $N_f:=\langle 0|\hat A_0^{\dagger}\hat A_0|0\rangle$
is the number of created zero mode gravitons
and we assumed that $N_f\gg 1$.
In deriving Eq.~(\ref{cP--Nf}) we used the commutation relation
$\left[\hat A_0, \hat A_0^{\dagger}\right]=1$
and the expression of the zero mode function~(\ref{mink_zeromode}).
The final power spectrum is then given by
\begin{eqnarray}
\cP(k):=\frac{4\pi k^3}{(2\pi)^3}\frac{2}{(M_5)^3}\cdot
\frac{1}{k\ell}N_f
=\frac{k^2}{\pi^2M_{{\rm Pl}}^2}N_f.
\label{def_ps}
\end{eqnarray}
Note here that by multiplying the factor $2/(M_5)^3$ we obtained
the squared amplitude of the metric fluctuation, rather than that of
the canonically normalized field [see the definition~(\ref{fourier})].

In fact, particle production is suppressed, $N_f\lesssim 1$,
for $k\gtrsim k_i:=a_iH_i$ where $a_i$ is the scale factor
at the end of inflation.
In the present paper we are interested in the modes
that exits the horizon during inflation, so that
the relevant wavenumber is smaller than $k_i$.

The operator $\hat A_0$ can be projected out
by making use of the Wronskian relations.
Noting that the Wronskian is constant in time,
we have
\begin{eqnarray}
\hat A_0&=&(\hat\phi\cdot \varphi_0)_f=(\hat\phi\cdot\Phi)
\nonumber\\
&=&(\phi_0\cdot\Phi)_i\hat a_0
+\int d\nu(\phi_{\nu}\cdot\Phi)_i\hat a_{\nu}+~\mbox{h.c.},
\end{eqnarray}
where $\Phi$ is a solution of the Klein-Gordon equation~(\ref{KGinTX})
whose final configuration is the zero mode function $\varphi_0$
in the Minkowski phase,
and
subscript $f$ and $i$ denote the quantities
on the final and initial hypersurfaces, respectively.
It is clear that final zero mode gravitons
are created from the vacuum fluctuations
both in the initial zero mode and in the KK modes:
\begin{eqnarray}
N_f=|(\phi_0^*\cdot \Phi)_i|^2+\int d\nu|(\phi_{\nu}^*\cdot \Phi)_i|^2.
\end{eqnarray}
Correspondingly, the power spectrum~(\ref{def_ps})
can be written as a sum of the two contributions:
\begin{eqnarray}
\cP=\cP_0+\cP_{{\rm KK}},
\end{eqnarray}
where
\begin{eqnarray}
\cP_0&:=&\frac{k^2}{\pi^2M_{{\rm Pl}}^2}|(\phi_0^*\cdot \Phi)_i|^2,
\\
\cP_{{\rm KK}}&:=&\frac{k^2}{\pi^2M_{{\rm Pl}}^2}\int d\nu|(\phi_{\nu}^*\cdot \Phi)_i|^2.
\end{eqnarray}
Thus what we need to do is
to solve the backward evolution of the field $\Phi$
and to evaluate the Wronskian on the initial hypersurface
in the de Sitter phase,
which can be done numerically~\cite{Kobayashi:2005jx, Kobayashi:2005dd}.

%~\cite{Kobayashi:2005jx, Kobayashi:2005dd}

 %\cite{Tanaka:2004ig, Kobayashi:2004wy}

%\cite{Kobayashi:2003cn}
  %``Primordial gravitational waves in inflationary braneworld,''
  %\cite{Kobayashi:2005jx}
  %``Quantum-mechanical generation of gravitational waves in braneworld,''
 %\cite{Tanaka:2004ig}
  %``AdS/CFT correspondence in a Friedmann-Lemaitre-Robertson-Walker brane,''
  %%CITATION = GR-QC 0402068;%%
%\cite{Kobayashi:2004wy}
  %``Leading order corrections to the cosmological evolution of tensor
  %perturbations in braneworld,''
%\cite{Hiramatsu:2003iz}
  %``Evolution of gravitational waves from inflationary brane-world:  Numerical
  %study of high-energy effects,''
  %%CITATION = HEP-TH 0308072;%%
%\cite{Hiramatsu:2004aa}
  %``Evolution of gravitational waves in the high-energy regime of brane-world
  %cosmology,''
  %%CITATION = HEP-TH 0410247;%%
%\cite{Ichiki:2003hf}
  %``Causal structure and gravitational waves in brane world cosmology,''
%\cite{Ichiki:2004sx}
  %``Stochastic gravitational wave background in brane world cosmology,''
%\cite{Kobayashi:2005dd}
  %``The spectrum of gravitational waves in Randall-Sundrum braneworld
  %cosmology,''
%\cite{Hiramatsu:2006bd}
  %``High-energy effects on the spectrum of inflationary gravitational wave
  %background in braneworld cosmology,''

\section{Inflationary gravitational waves with initial KK fluctuations}

Let $\delta_T$ be the primordial amplitude of gravitational waves
from de Sitter inflation, which is given by~\cite{Langlois:2000ns}
\begin{eqnarray}
\delta_T^2=\frac{2C^2(\ell H_i)}{\mpl^2}\left(\frac{H_i}{2\pi}\right)^2,
\end{eqnarray}
where $H_i$ is the Hubble parameter during inflation and
the function $C(\ell H_i)$ is related to the
normalization of the zero mode (see Appendix).
After the end of inflation
the brane universe is
dominated by a perfect fluid whose equation of state parameter
is $w$,
and the FRW phase is connected to the Minkowski phase
when the energy scale of the universe becomes sufficiently
low: $H= H_0\ll \ell^{-1}$.
For the modes that reenter the horizon during the low energy regime,
namely, for the long wavelength modes with $k\ll k_*$
where
\begin{eqnarray}
k_*:= a_* H_* = a_*/\ell,
\end{eqnarray}
corrections to their evolution are very small,
suppressed by $\ell^2$ and $\ell^2\ln \ell$~\cite{Tanaka:2004ig, Kobayashi:2004wy};
the mode mixing effect is inefficient and
gravity on the brane is basically described by
four-dimensional general relativity.
Hence, for such modes the spectrum of gravitational waves
is expected to have the same spectrum as the four-dimensional one:
\begin{eqnarray}
\cP=\frac{\delta^2_T}{2}\left(\frac{k}{k_0}\right)^{n_{{\rm 4D}}},
\quad
(k_0<k\ll k_*),
\end{eqnarray}
with
\begin{eqnarray}
n_{{\rm 4D}} = -\frac{4}{1+3w},
\end{eqnarray}
where $k_0$ is the wavenumber associated with
the horizon scale at the end of the FRW phase: $k_0:=a_0H_0$.
For the relatively short wavelength modes with $k\gtrsim k_*$,
three non-trivial things should be considered:
(i) the modified background expansion rate [Eq.~(\ref{friedmanneq})],
(ii) the excitation of KK modes relevant at high energies,
and (iii) the initial quantum fluctuations in KK modes $\cP_{{\rm KK}}$.
If the gravitational waves propagating in the bulk are neglected,
namely, if the effects (ii) and (iii) are neglected,
one would obtain an enhanced spectrum for $k\gtrsim k_*$,
\begin{eqnarray}
\cP_{\uparrow} \simeq \frac{\delta_T^2}{2}\left(\frac{k_*}{k_0}\right)^{-2}
\left(\frac{k}{k_*}\right)^{n_{\uparrow}},
\label{uparrow1}
\end{eqnarray}
with
\begin{eqnarray}
n_{\uparrow}=-\frac{2}{2+3w}.\label{uparrow2}
\end{eqnarray}
However, this evaluation will not be correct because of
efficient mode mixing at high energies.\footnote{There
are also braneworld models where one arrives at
an interesting conclusion about cosmological gravitational waves
{\em without} mode mixing~\cite{Cavaglia:2005id, DeRisi:2006pz}.
}
The true spectrum will be somewhat different from the four-dimensional one
and differed also from Eqs.~(\ref{uparrow1}) and~(\ref{uparrow2}).
We would like to investigate how the balance of the above three things
affects the spectrum.

In Refs.~\cite{Hiramatsu:2004aa, Ichiki:2004sx, Hiramatsu:2006bd} the effects of
(i) the modified Friedmann equation and (ii) the KK mode excitation
have been addressed using the numerical formulations very different
from ours.\footnote{The formulation by
Hiramatsu \textit{et al.}~\cite{Hiramatsu:2004aa}
is based on the Poincar\'{e} coordinates, while
Ichiki and Nakamura~\cite{Ichiki:2004sx} used a single null coordinate.
The two results do not agree with each other for some unclear reason.
However, the Wronskian approach~\cite{Kobayashi:2005dd} gives
the result consistent with Ref.~\cite{Hiramatsu:2004aa}.
}
The authors of Refs.~\cite{Hiramatsu:2004aa, Ichiki:2004sx, Hiramatsu:2006bd}
focused on the classical evolution of perturbations,
while the initial conditions they adopt
are na\"{i}ve,
neglecting the initial abundance of KK fluctuations.
In our Wronskian formulation,
the initial conditions are set quantum-mechanically
and hence are the most plausible.
It has been (incompletely)
shown that the effect of the initial quantum fluctuations
in the KK modes is subdominant relative to the other two effects
in the radiation-dominated phase ($w=1/3$)~\cite{Kobayashi:2005dd}.
In that case, however, the dominant two effects
cancel each other and consequently we have the same spectrum
as in the conventional four-dimensional universe.

\begin{figure}[tb]
  \begin{center}
    \includegraphics[keepaspectratio=true,height=50mm]{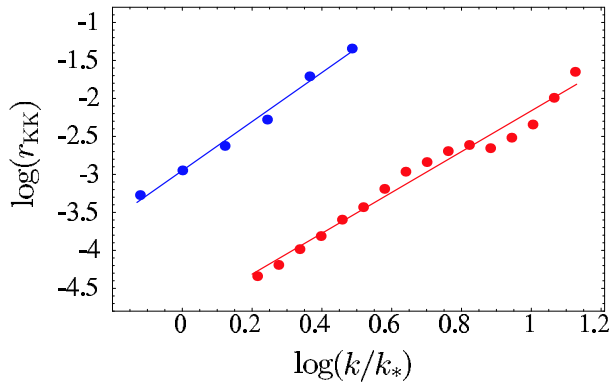}
  \end{center}
  \caption{(color online). Contribution of initial Kaluza-Klein fluctuations.
  The equation of state parameter is $w=1/3$ and
  the inflationary energy scale is given by $\ell H_i=100$ (lower line).
  $r_{{\rm KK}}$ is plotted also for $\ell H_i=10$ (upper line).
  (The previous estimate in Ref.~\cite{Kobayashi:2005dd} should
  be compared with the lower line.)}%
  \label{fig:rKK_rad.eps}
\end{figure}

In the previous paper~\cite{Kobayashi:2005dd}
the authors failed in the precise evaluation
of the fraction coming from the initial KK modes,
\begin{eqnarray}
r_{{\rm KK}}(k) := \frac{\cP_{{\rm KK}}}{\cP_0+\cP_{{\rm KK}}},
\label{rkk}
\end{eqnarray}
due to the limited number of grids in the extra direction.
Now using the new coordinates $T$ and $X$,
we are able to obtain much more precise values of $r_{{\rm KK}}$.
Our refined result is shown in Fig.~\ref{fig:rKK_rad.eps}.
This is different from the previous estimate by a factor
of two or so.

\begin{figure}[tb]
  \begin{center}
    \includegraphics[keepaspectratio=true,height=50mm]{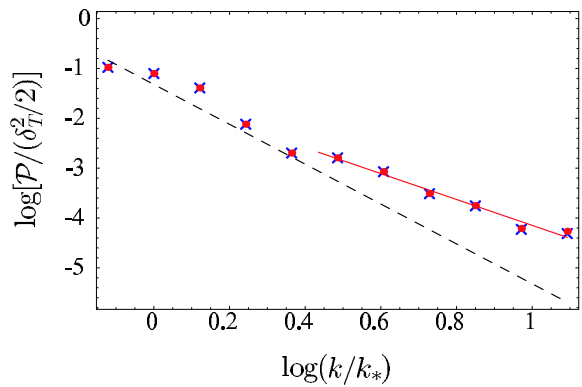}
  \end{center}
  \caption{(color online). Spectrum of gravitational waves from inflation with $\ell H_i=100$,
  followed by a FRW phase with $w=0$.
  The total power spectrum is shown by red circles, while
  blue crosses represent the contribution only from the initial zero mode.
  Dashed line indicates the four-dimensional result.}%
  \label{fig:Pw0.eps}
\end{figure}
\begin{figure}[tb]
  \begin{center}
    \includegraphics[keepaspectratio=true,height=50mm]{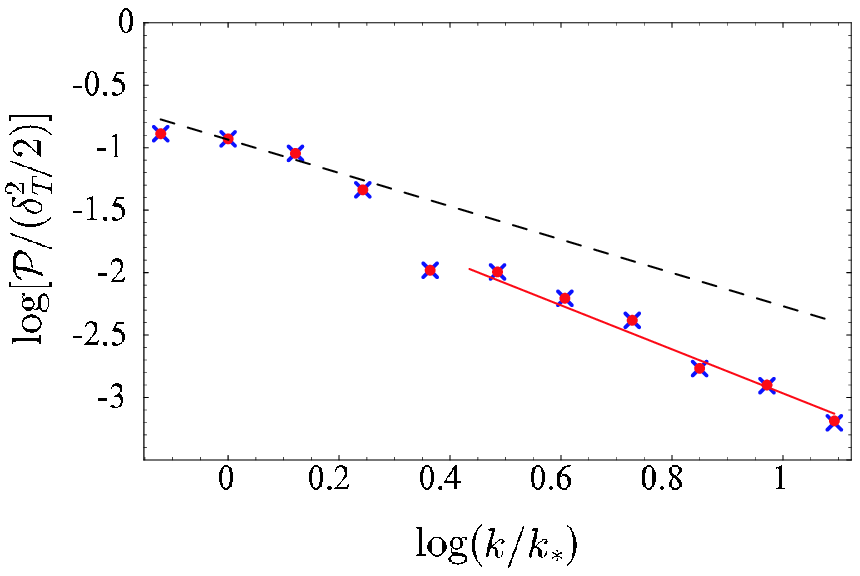}
  \end{center}
  \caption{(color online). Same as Fig.~\ref{fig:Pw0.eps}, but $w=2/3$.}%
  \label{fig:Pw2-3.eps}
\end{figure}
\begin{figure}[tb]
  \begin{center}
    \includegraphics[keepaspectratio=true,height=50mm]{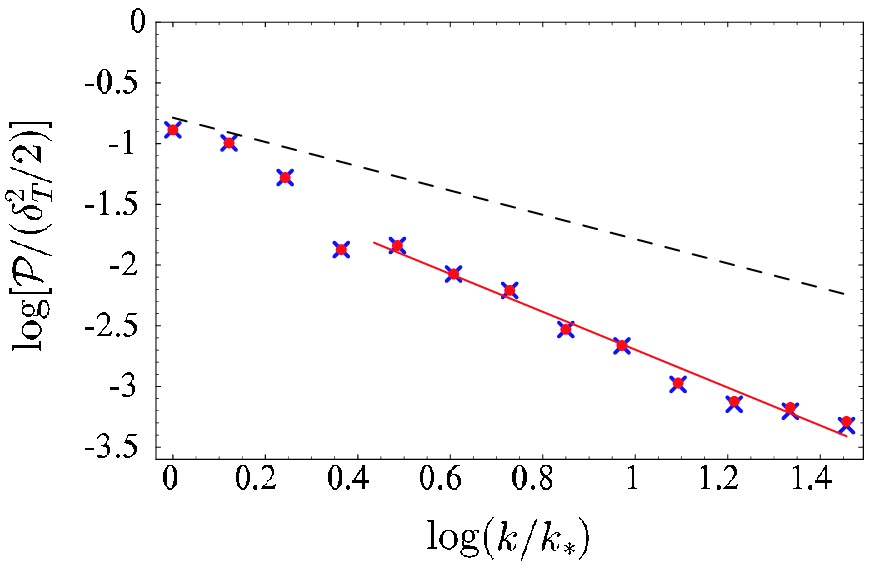}
  \end{center}
  \caption{(color online). Same as Fig.~\ref{fig:Pw0.eps}, but $w=1$.}%
  \label{fig:Pw1.eps}
\end{figure}

To make it clear how the spectrum behaves in a more general situation,
let us move on to the models with the equation of state parameter
other than $1/3$.
We have performed numerical calculations
for three different cases: $w=0, 2/3$, and 1.
The results are shown in Figs.~\ref{fig:Pw0.eps}--\ref{fig:Pw1.eps}.
We find that {\em in all the cases 
the modification to the expansion rate and the KK mode excitation
are the dominant effects, and the initial abundance of
KK fluctuations gives a subdominant contribution.}
In contrast to the radiation-dominated model,
the two dominant effects do not cancel each other,
yielding the spectrum different from the four-dimensional one
for short wavelength modes.
Assuming that the spectrum has the form
\begin{eqnarray}
\cP \propto k^n, \quad k>\alpha k_*,
\end{eqnarray}
where we take $\alpha\sim 3$,
we obtain
\begin{eqnarray}
n\simeq\left\{
  \begin{array}{l}
       -2.6,\quad(w=0)\\
       -1.8,\quad(w=2/3)\\
       -1.6,\quad(w=1).\\
  \end{array}
\right.
\end{eqnarray}
From this result and the previous one ($n\simeq -2$ for $w=1/3$),
we may deduce that the spectral tilt for short wavelength modes
is given generally by
\begin{eqnarray}
n \simeq - \frac{5+3w}{2+3w}.
\label{spectral_index}
\end{eqnarray}
This agrees with the recent numerical result
by Hiramatsu~\cite{Hiramatsu:2006bd}
because of the insensitivity of the final result
to the initial KK mode contamination.
For the modes with $k\lesssim k_*$,
we can confirm that the spectrum is approximately given by
the four-dimensional one in all the three cases.

\begin{figure}[tb]
  \begin{center}
    \includegraphics[keepaspectratio=true,height=50mm]{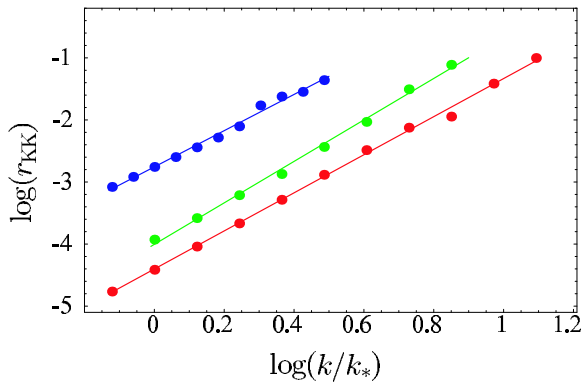}
  \end{center}
  \caption{(color online). Fraction $r_{{\rm KK}}$ for different inflationary energy scales,
  showing (from up to bottom) $\ell H_i=10, 50,$ and $100$.
  The equation of state parameter is given by $w=0$.}%
  \label{fig:rKKw0.eps}
\end{figure}

\begin{figure}[tb]
  \begin{center}
    \includegraphics[keepaspectratio=true,height=50mm]{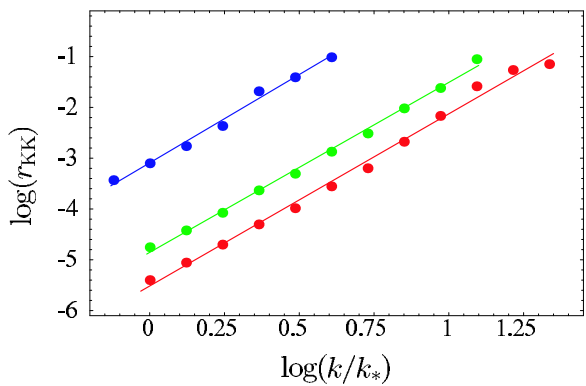}
  \end{center}
  \caption{(color online). Same as Fig.~\ref{fig:rKKw0.eps}, but $w=1$.}%
  \label{fig:rKKw1.eps}
\end{figure}

Now we take a closer look at the contribution
from the initial KK fluctuations.
The fraction $r_{{\rm KK}}$ is plotted for
various values of the inflationary Hubble parameter $H_i$
in Figs.~\ref{fig:rKKw0.eps} and~\ref{fig:rKKw1.eps}
(and also in Fig.~\ref{fig:rKK_rad.eps}).
While the equation of state parameter is different
($w=0$ for Fig.~\ref{fig:rKKw0.eps} and $w=1$ for Fig.~\ref{fig:rKKw1.eps}),
the plots share the same feature.
For fixed $k$, $r_{{\rm KK}}$ increases with decreasing $\ell H_i$,
but, interestingly, the maximum of $r_{{\rm KK}}$ is given by $\sim 0.1$
irrespective of the inflationary energy scale $\ell H_i$. 
We find that
this behavior can be summarized in the following universal relation:
\begin{eqnarray}
r_{{\rm KK}}(k)\simeq r_{{\rm max}}\left(\frac{k}{k_i}\right)^{\beta},
\label{rKK_relation}
\end{eqnarray}
with
\begin{eqnarray}
r_{{\rm max}} &<& {\cal O}(1),\\
\beta &\approx& 3,
\label{beta<3}
\end{eqnarray}
and note that $k_i$ is given by
\begin{eqnarray}
k_i=a_iH_i \simeq (\ell H_i)^{(2+3w)/(3+3w)}\times k_*.
\end{eqnarray}
Our numerical analysis indicates that the formula~(\ref{rKK_relation}) holds
as long as the inflationary stage lies in the high energy regime.
For low energy inflation ($\ell H_i\ll 1$) $r_{{\rm KK}}$
is further suppressed, and
we confirmed that by computing $r_{{\rm KK}}$
for a model with $\ell H_i=0.2$ and $w=1/3$, showing that
$r_{{\rm KK}}\sim 10^{-3}$ at $k\sim a_iH_i$.

\section{Conclusions}

In this paper, we have investigated the spectrum of
gravitational waves generated from inflation on the brane.
Our analysis generalize the previous work~\cite{Kobayashi:2005dd}
in that the present background cosmological model allows for
the post-inflationary stage characterized by the equation of state parameter $w$
other than $1/3$.
We have paid a particular attention to the effect of
the initial condition: the configuration of the perturbations during inflation
includes vacuum fluctuations in Kaluza-Klein modes.
This was made possible  by the use of the Wronskian
formulation~\cite{Kobayashi:2005jx, Kobayashi:2005dd},
which is a distinctive point compared to the other related works.

Our numerical analysis
for the various inflationary energy scale and equation of state parameter
showed that
the contribution of the initial KK fluctuations $r_{{\rm KK}}(k)$,
defined in Eq.~(\ref{rkk}), always behaves as Eq.~(\ref{rKK_relation}),
implying that \textit{the initial KK fluctuations give rise to the subdominant effect
on the spectrum}.
This is the main conclusion of the paper, being true
in the more general setup than the previous work~\cite{Kobayashi:2005dd}.
The present calculation covers the limited range of
the inflationary energy scale (up to $\ell H_i = 100$),
but we believe
the results to be sufficient for concluding that
also for higher energy scales,
$r_{{\rm KK}}$ is smaller than order of unity
even at the smallest wavelength concerned ($k\sim k_i$).
While the effect of the initial KK fluctuations
can be safely neglected,
the enhancement effect due to the modified Friedmann equation
and the damping effect due to the excitation of KK modes
are dominantly work, yielding the spectral tilt
$n\simeq -(5+3w)/(2+3w)$ for short wavelength modes.
This can be a observational signature of the braneworld model
with nonstandard (i.e., $w\neq 1/3$) cosmological history.

%This is in general different from the four-dimensional prediction,
%except for the radiation case~\cite{Hiramatsu:2006bd}.

All the results in this paper are given numerically,
but surely they will be great help for understanding
the evolution of gravitational wave perturbations in the braneworld.
Now the clues are in order:
the spectral index is given by Eq.~(\ref{spectral_index})
and it is insensitive to the contamination of the initial KK fluctuations,
with their effect expressed in detail by Eqs.~(\ref{rKK_relation})--(\ref{beta<3}).
We will hopefully provide some analytic arguments
complementary to the present numerical results
in a future publication.

\acknowledgments
I would like to thank Takahiro Tanaka
for useful comments.
I am supported by the JSPS under Contract No.~01642.
%%%%%%%%%%%%%%%%%%%%%%%%%%%%%%%%%%%%%%%%%%%%%%%%%%%%%%%%%%%%%%%%%%%%%%
\appendix
\section{Mode functions}

In this appendix, we present an explicit form of
the mode functions in the Minkowski and de Sitter braneworlds.

In the Minkowski braneworld, the normalized zero mode function is given by
\begin{eqnarray}
\varphi_0(t)=\frac{1}{\sqrt{2k\ell}}e^{-ikt},
\label{mink_zeromode}
\end{eqnarray}
while the normalized KK mode function is
\begin{eqnarray}
\varphi_m(t,z)=\frac{1}{\sqrt{2\omega \ell^3}}e^{-i\omega t}u_m(z),
\end{eqnarray}
with
\begin{eqnarray}
u_m(z):=z^2\sqrt{\frac{m}{2}}
\frac{Y_1(m\ell)J_2(mz)-J_1(m\ell)Y_2(mz)}
{\sqrt{[Y_1(m\ell)]^2+[J_1(m\ell)]^2}},
\end{eqnarray}
and
\begin{eqnarray}
\omega=\sqrt{k^2+m^2}.
\end{eqnarray}

In the de Sitter braneworld we introduce
new coordinates $(\eta, \xi)$,
which is related to $(t, z)$ as
\begin{eqnarray}
t=\eta\cosh\xi+t_0,\quad z=-\eta\sinh\xi,
\end{eqnarray}
where $t_0$ is an arbitrary constant.
In $(\eta,\xi)$ frame
the de Sitter brane is located at a fixed coordinate position $\xi=\xi_b=$ constant,
and the Hubble parameter on the brane is given by $H_i=\ell^{-1}\sinh\xi_b$.
The normalized zero mode is
\begin{eqnarray}
\phi_0(\eta)=C(\ell H_i)\cdot
\frac{H_i}{\sqrt{2k\ell}}\left(
\eta-\frac{i}{k}\right) e^{-ik\eta},
\end{eqnarray}
with
\begin{eqnarray}
C(x):=\left[
\sqrt{1+x^2}+x^2\ln\left(\frac{x}{1+\sqrt{1+x^2}}\right)\right]^{-1/2},
\end{eqnarray}
and the KK mode functions are found in the form of
$\phi_{\nu}(\eta,\xi)=\psi_{\nu}(\eta)\chi_{\nu}(\xi)$,
where
\begin{eqnarray}
\psi_{\nu}(\eta) &=& \frac{\sqrt{\pi}}{2}\ell^{-3/2}
e^{-\pi\nu/2}(-\eta)^{3/2}H^{(1)}_{i\nu}(-k\eta),
\\
\chi_{\nu}(\xi) &=& C_1(\sinh\xi)^2
\left[
P^{-2}_{-1/2+i\nu}(\cosh\xi)\right.
\nonumber\\
&&\quad\left.-C_2
Q^{-2}_{-1/2+i\nu}(\cosh\xi) \right],
\end{eqnarray}
with
\begin{eqnarray}
C_1&=&\left[\left| \frac{\Gamma(i\nu)}{\Gamma(5/2+i\nu)}\right|^2\right.
\nonumber\\
&&\hspace{-10mm}
\left.+\left| \frac{\Gamma(-i\nu)}{\Gamma(5/2-i\nu)}
-\pi C_2\frac{\Gamma(i\nu-3/2)}{\Gamma(1+i\nu)}\right|^2\right]^{-1/2},
\\
C_2&=&
\frac{P^{-1}_{-1/2+i\nu}(\cosh\xi_b)}{Q^{-1}_{-1/2+i\nu}(\cosh\xi_b)}.
\end{eqnarray}

%%%%%%%%%%%%%%%%%%%%%%%%%%%%%%%%%%%%%%%%%%%%%%%%%%%%%%%%%%%%%%%%%%%%%%

%---------   References   ---------%

%---------   References   ---------%

\end{document}